\documentstyle[prb,twocolumn,aps, epsf]{revtex}
\begin{document}
\twocolumn[\hsize\textwidth\columnwidth\hsize\csname @twocolumnfalse\endcsname

\title{Photoluminescence Detected Doublet Structure\\
in the Integer and Fractional Quantum Hall Regime}
\author{F. M.\, Munteanu$^{*,\dag}$, Yongmin\, Kim$^*$, C. H.\,
Perry$^{*,\dag}$, D. G.\, Rickel$^*$,\\
J. A.\, Simmons$^{\ddag}$, and J. L.\, Reno$^{\ddag}$}
\address{$^*$National High Magnetic Field Laboratory-Los Alamos
National Laboratory\\
Los Alamos, NM 87545\\
$^{\dag}$Department of Physics, Northeastern University, Boston, MA
02115\\ $^{\ddag}$Sandia National Laboratory, Albuquerque, NM 87185}
\date{\today} \maketitle
\begin{abstract}
We present here the results of polarized magneto-photoluminescence
measurements on a high mobility single-heterojunction. The presence of a
doublet structure over a large magnetic field range (2$>$$\nu$$>$1/6) is
interpreted as possible evidence for the existence of a magneto-roton
minima of the charged density waves. This is understood as an indication of
strong electronic correlation even in the case of the IQHE limit.
\end{abstract}
\pacs{78.66.Fd, 73.40.Hm, 78.20.Ls}

]\narrowtext

The use of magneto-photoluminescence (MPL) spectroscopy to study the
integer (IQHE) and fractional quantum Hall effects (FQHE) has
attracted considerable interest in recent years.  In the case of the
IQHE, the appearance of the gap in the spectrum at the Fermi energy is
caused by the quantization of the magnetic Landau levels or the
quantization of the spin energy levels and is essentially a single
particle phenomenon.  In the case of FQHE, the appearance of the gap
is understood as a result of condensation of the 2 DEG into an
incompressible quantum liquid (IQL)\cite{1} due to strong
electron-electron correlations which takes place at fractional filling
factors $\nu$=p/q$<$1.

One of the most significant phenomenon that occurs in the FQHE regime
is the emergence of low-lying charge-density (CD) waves that display
characteristic magneto-roton (MR) minima at a wave vector close to the
inverse magnetic length.  At large wave vector, neutral CD waves
excitations consist of pairs of fractionally charge quasiparticles
that are associated with the energy gaps of the IQL. More recently, it
was suggested that the strong correlations between electrons are
important in explaining the structure observed in photoluminescence at
$\nu$=1 and this state is considered to be a strongly correlated
state,\cite{2} similar to the incompressible states at fractional
filling factors.  It was shown that the system always has a gap, even
when the single-particle gap vanishes (i.e. when g=0), as a result of
electron-electron repulsion.  It has also been shown\cite{3} that, in
that case when the electron Zeeman energy is large, the low energy
states at $\nu$=1 are the excitonic states in which the
spin-$\uparrow$ lowest Landau level is filled and the valence band
hole binds with a single spin-$\downarrow$ electron to form an
exciton.  In analyzing the PL spectra obtained from a 2 DEG in the
FQHE regime, the appearance of a doublet structure in the
photoluminescence (PL) spectrum\cite{4,5,6,7} around $\nu$=2/3 was
interpreted by Apalkov and Rashba\cite{8} as an evidence for the
appearance of an indirect, single MR transition from an extensive area
in k space for {\it k}$\ell$$_{B}$$\cong$1 (where $\ell$$_{B}$ is the
magnetic length).  The formation of the rotons is due to the
reduction, at large wave vectors, of the excitonic binding energy
between the electron and the hole.\cite{9,10,11} The MR minimum is a
precursor to the gap collapse associated with the Wigner crystal
instability \cite{12} and is connected with the excitonic attraction
between fractionally charged quasiparticles.  The calculations
\cite{8} reported for the case of the filling factor $\nu$=1/3 showed
that the dispersion of the excitons is strongly suppressed by their
coupling to the IQL and the lowest branch of the exciton spectrum
passes completely below the MR spectrum.  Another important result of
note is the fact that, even if its theoretical prediction is based on
calculations performed in the case of fractional filling factor
$\nu$=1/3, the existence of the MR peak was observed in the MPL, also
at filling factors that do not involve the FQHE.\cite{4,5,6,7} In
addition, evidence of excitonic binding and roton minima formation was
found by Pinczuk {\it et.al.} for the filling factors $\nu$$>$1
\cite{13} and $\nu$=1/3 \cite{14} from inelastic light scattering
studies performed on the 2DEG formed in high-mobility GaAs structures.
Karrai {\it et.al.} \cite{15} analyzed the magneto-transmission spectra
obtained from a quasi-three-dimensional electron system subjected to a
parallel magnetic field and found evidence for the existence of a
MR excitation for a wide range of magnetic fields.

In this report, we present the results of MPL measurements of a MBE
grown high quality GaAs/Al$_{0.3}$Ga$_{0.7}$As single heterojunction
(SHJ) with a dark electron density of 1.2$\times$10$^{11}cm^{-2}$ and
a mobility higher than 3$\times$10$^{6}cm^{2}$/Vs.  In these
experiments, during constant laser illumination, the 2DEG density
increased to 2.1$\times$10$^{11}cm^{-2}$.  The experimental layout for
the PL measurements has been described previously.\cite{16} Using a
quasi-continuous magnet, the field was varied from 0 to 60 T,
while the temperature was changed from 1.5K to 450mK. The polarized
spectra that we obtained showed the appearance of a doublet at filling
factors $\nu$$>$3/2 and the persistence of this effect to the highest
magnetic fields utilized.  In Fig.  1 we show the unpolarized spectra
obtained at a temperature of 1.5K at the filling factor $\nu$=2, 3/2
and 1.  The E0-hh peak that appears at $\nu$=5 
\begin{figure}[tb] 
    \caption{Unpolarized MPL spectra at 1.5K for three different filling
factors. The appearance of the doublet structure is resolved at $\nu$=3/2.
The  inset shows the difference ($\Delta$) between the
magneto-roton (MR) and the exciton (X) energies as a function of magnetic
field to 58T.}
    \end{figure}
(B=1.82T) shows a
splitting for 2$>$$\nu$$>$1 and this is clearly shown in the spectra
at $\nu$=3/2.  This splitting, once formed, is present for the whole
range of magnetic fields examined.  We believe that the lower energy
peak of the doublet is associated with neutral exciton (X) transition,
while the higher energy peak may be evidence of a MR
transition.  The difference in the energies ($\Delta$) between these
two peaks as a function of magnetic field is shown in the inset of the
Fig.1.  It can be seen that $\Delta$ has a sudden increase in the
region of $\nu$=1/2 (16-19T) and then saturates for fields higher than
30T, a behavior similar to that reported by Heiman {\it et.al.}
\cite{4,5} in their MR studies.  The value of the
separation (0.4-1.2 meV) is close to the FQHE quasiparticle-quasihole
separation gap energy (about 1 meV).  It does not scale as the
magnetic energy (B$^{1/2}$) but rather follows an almost linear
behavior in the range 1$>$$\nu$$>$1/3.

Cooper and Chklovskii \cite{3} noted that in the situation where the
valence-band hole is close to the electron gas compared with the
electron-electron spacing, the most important initial states are the
excitonic states, in which the Landau level of spin-$\uparrow$
electrons is fully occupied, and the valence-band hole binds with a
spin-$\downarrow$ electron to form an exciton.  The excitonic states
will compete for the initial ground state of the system with the
"free-hole" states in which the photoexcited electron fills the vacant
spin-$\uparrow$ state and the hole occupies the lowest Landau level
single particle state.  As the filling factor is swept from $\nu$$>$1 to
$\nu$$<1$, the form of the initial state contributing to the PL signal
is believed to undergo a transition from an excitonic state to a
"free-hole" state and a red shift in the luminescence line should be
expected.  The fact that in the case of our sample, there is no
discontinuity in the energy line is an indication that there is no
change in the nature of the initial excitonic state.  The lack of the
transition to a "free-hole" state in our case is also 

\begin{figure}[tb] 
    \caption{The evolution of the integrated intensities of the 
magneto-roton (MR) and exciton (X) peaks with magnetic field at 1.5K. 
The excitonic line (X) shows distinctive minima at $\nu$=1, 2/3, 5/9, 
4/11, 1/3.  The intensity behavior of the MR line shows a deep minimum 
at $\nu$=8/11 and local maxima at $\nu$=2/5 and 1/3.}
    \end{figure}
confirmed by the
absence of the additional peak in the LCP polarization as observed by
Plentz {\it et.al.}.\cite{17}

	The evolution of the intensities of the two peaks is shown in Fig.
2. The excitonic line shows distinctive minima at $\nu$=1, 2/3, 5/9, 4/11, 1/3.
Similar behavior has been reported by Turberfield {\it et.al.} \cite{6} and
can be related to the localization of the electrons in these states
concomitant with a reduction of the screening factor. The intensity
behavior of the line labeled MR shows a deep minimum at $\nu$=8/11 and local
maxima at $\nu$=2/5 and 1/3 (1/3 fractional hierarchy). A similar transfer
of intensity from the lower energy line to the higher energy one at $\nu$=1/3
was previously reported by Heiman {\it et.al.} \cite{4,5} and may be caused
by an enhancement of the CD wave as a result of a reduction of 
screening.\\

Figure 3 is intentionally removed due to its size.

\begin{figure}[tb] 
    \caption{Right circularly polarized (RCP) spectra at 1.5K at low and 
high magnetic fields.  The spectra taken at low magnetic fields show a 
decrease of the intensity of the lower energy peak with magnetic field 
as a result of depopulation of the spin-$\downarrow$ electronic level.  
At high magnetic field, the intensities of the two lines are almost 
the same.}
    \end{figure}

\begin{figure}[tb] 
    \caption{The ratio of the magneto-roton (MR) and exciton (X) 
intensities for right (RCP) and left (LCP) circularly polarized 
spectra as a function of magnetic field.}
    \end{figure}
    
    Fig.  3 shows the right circularly polarized (RCP) spectra at the same
temperature of 1.5K at low and high magnetic fields.  It can be seen
that the spectra taken at low magnetic fields show a decrease of the
intensity of the lower energy peak with magnetic field as a result of
depopulation of the spin-$\downarrow$ electronic level, a behavior
that is expected in the case of the neutral exciton.  At large
magnetic fields, the intensities of the two peaks are almost the
same as a result of low filling factor. 
 Fig.  4 shows the ratios of
the intensities of the MR and excitonic peaks for the right
(RCP) and left (LCP) circular polarizations.  In the LCP case this
ratio is less than one as a result of a higher intensity of the
excitonic peak compared to the RCP case.  The difference in the
energies (not shown) of the two peaks as a function of magnetic field
is almost the same for both polarizations; like the unpolarized data,
this difference does not scale as B$^{1/2}$ as may be
expected.\cite{8} Haldane and Rezay\cite{8} estimated a separation
value $\Delta$=0.075e$^{2}$/$\epsilon$$\ell$$_{B}$ (where
$\epsilon$=12.8 for GaAs) which is larger than the value that we
measured for the magnetic fields in excess of 15T ($\nu$=0.6).  The
evolution of the energies with the magnetic field in both
polarizations at the filling factor $\nu$=1 does not show any
significant blue or red shift.

	In the case of narrow quantum wells it
has been shown, both theoretically and experimentally, that the
correlation hole of the hole term is the primary mechanism that
generates the blue shifted energy associated with the IQHE.\cite{19}
In the case of wider quantum wells, a red shift in energy at $\nu$=1
indicates that the screening is reduced and the electron-hole Coulomb
interaction (vertex correction) is enhanced.\cite{19} The fact that no
significant 
\begin{figure}[tb] 
    \caption{Right (RCP) and left (LCP) circularly polarized spectra taken at
two different temperatures,1.5K and 450 mK, for filling factors (a)
$\nu$=1/5  and (b) $\nu$=1/3 . All the intensities have been normalized
with respect to the zero field spectra.}
    \end{figure}
energy shift is seen in both RCP and LCP spectra at
filling factor $\nu$=1 seems to indicate that the cancellation of the
screened exchange and Coulomb hole terms for the electrons is close to
exact, while the vertex correction term and correlation hole of the
hole term cancel each other to a very good degree.  This can be
interpreted as an evidence of an incomplete cancellation of screening
for this filling factor.  Because of the finite temperatures, we
expect that some of the low energy excited states will also be
populated.  These states will consist of both finite momentum exciton
states and of long wavelength spin wave excitations of the system.
They will be seen as fluctuations in the overall polarization of the
system and will lead to a mixing between the two circular
polarizations.  For this reason, a degree of uncertainty in the
measured ratio of the intensities is possible.  For the LCP spectra,
the doublet is resolved at magnetic fields higher than in the RCP
spectra.  We believe that this may be caused by a broadening of the
excitonic transition line as a result of the presence of the spin
waves in the LCP spectra.  This appears as a result of the
recombination of one of the valence-band holes with a spin-$\uparrow$
electron, a process that leaves a spin reversal in the final state.

	Fig. 5 a and b show the MPL spectra in both LCP and RCP
polarizations for two temperatures (1.5K and 450mK) at two filling factors
$\nu$=1/5 and 1/3. All the intensities are normalized with respect to the
zero field spectra. It can be seen that, by decreasing the temperature, the
higher energy peak is the one that becomes stronger, ruling out the
possibility of it being generated by the changes in the population of
electrons and photo-excited holes. A similar behavior was reported by
Heiman{\it et.al.} \cite{4} and was explained by Apalkov and Rashba
\cite{8} as being due to the proximity between the electron and hole
confinement planes. This proximity can also be explained by a process
similar to the one described by Kim {\it et.al.} \cite{20}, namely the
reduction in screening between electrons, and is the main reason for the
existence of the excitonic states at $\nu$=1. In Fig. 4a the intensity of
the excitonic peak in the LCP spectra is clearly larger than that observed
in the RCP spectra. The poorer resolution of the two peaks in the LCP
spectra compared to the RCP spectra can be explained by the broadening of
the excitonic transition line as a result of the presence of the spin waves
in the LCP spectra.

	In conclusion, MPL  measurements have been performed on a high
quality GaAs/AlGaAs SHJ. The spectra showed a doublet structure over a
large range of the magnetic field (2$>$$\nu$$>$1/6) that we believe to be
evidence for the coexistence of the E0 excitonic excitations and
MR minima of the CD waves. Their presence over such a wide field
range and filling factor is understood as a confirmation of the strong
correlation effects among electrons in the integer and quantum Hall regime.

	The authors gratefully acknowledge the engineers and technicians at
NHMFL-LAPF in the operation of the 60T QC magnet. Work at NHMFL-LAPF is
supported by NSF Cooperative Agreement  DMR 9527035,  the Department of
Energy and the State of Florida. Work at Sandia National Laboratory is
supported by the Department of Energy.

\end{document}